\documentclass[12pt]{article}

\usepackage{times}
\pdfoutput=1
\newenvironment{sciabstract}{%
\begin{quote} \bf}
{\end{quote}}

\usepackage[latin1]{inputenc}  %
\usepackage[pdftex]{graphicx}

\usepackage{amssymb}
\usepackage{amsmath}

\def\ba#1\ea{\begin{align*}#1\end{align*}} %
\def\banum#1\eanum{\begin{align}#1\end{align}} %

\newcommand{\bi}{\begin{itemize}}
\newcommand{\ei}{\end{itemize}}
\newcommand{\be}{\begin{enumerate}}
\newcommand{\ee}{\end{enumerate}}
\newcommand{\bc}{\begin{center}}
\newcommand{\ec}{\end{center}}

\newcommand{\condon}{\, | \,}%
\newcommand{\R}{I\!\!R}

\newtheorem{propositionnn}{Proposition}

\newcommand{\STt}[3]{$t(#1) = #2$, $#3$} %

\usepackage{hyperref}
\usepackage{apacite}
\newcommand{\pressLink}[1]{ {\tiny \url{#1}}}
\title{Unconscious lie detection as an example of a widespread fallacy in the Neurosciences}

\author
{Volker H. Franz,$^{1\ast}$ Ulrike von Luxburg$^{2}$\\
\\
\normalsize{$^{1}$Department of Psychology, University of Hamburg,}\\
\normalsize{von Melle Park 5, 20146 Hamburg, Germany}\\
\normalsize{$^{2}$Department of Computer Science, University of Hamburg,}\\
\normalsize{Vogt--Koelln--Str. 30, 22527 Hamburg, Germany}\\
\\
\normalsize{$^\ast$To whom correspondence should be addressed; E-mail:  volker.franz@uni-hamburg.de.}
}

\date{}

\begin{document}

\maketitle

\begin{sciabstract}
  Neuroscientists frequently use a certain statistical reasoning to
  establish the existence of distinct neuronal processes in the
  brain. We show that this reasoning is flawed and that the large
  corresponding literature needs reconsideration. We illustrate the
  fallacy with a recent study that received an enormous press coverage
  because it concluded that humans detect deceit better if they use
  unconscious processes instead of conscious deliberations. The study
  was published under a new open--data policy that enabled us to reanalyze
  the data with more appropriate methods. We found that unconscious
  performance was close to chance -- just as the conscious
  performance. This illustrates the flaws of this widely used
  statistical reasoning, the benefits of open--data practices, and the
  need for careful reconsideration of studies using the same
  rationale.
\end{sciabstract}

\newpage 

\subsection*{Introduction}

\noindent Lie detection is of considerable importance to modern
society, in particular in connection with police investigations, court
proceedings, and security questions. For example, the U.S. government
invests large amounts of money for training ``behavior detection
officers'' to detect terrorists from their behavior at airports. These
programs have been criticized for being irrational
\cite{press_NYT_20140323} because scientific evidence suggests that
humans are only correct in approximately 54\% %
of lie--truth judgments \cite{Bond_DePaulo_06}. This is essentially
as good as flipping a coin. In this context, the recent lie detection
study \cite{tenBrinke_14} presents the surprising finding that
unconscious processes are much better in detecting liars than
conscious processes. Consequently, the study received enormous
attention\footnote{
Selected press coverage (retrieved Mar--May 2014): 
New York Times, Apr. 26 \pressLink{http://www.nytimes.com/2014/04/27/business/the-search-for-our-inner-lie-detectors.html}
Science Magazine, Apr 1 \pressLink{http://news.sciencemag.org/signal-noise/2014/03/spot-liar-trust-your-instinct}
BBC, Mar 29 \pressLink{http://www.bbc.com/news/health-26764866}
British Psychological Society, Mar 28 \pressLink{http://www.bps.org.uk/news/our-subconscious-mind-may-detect-liars}
S{\"u}ddeutsche Zeitung, Mar 27 \pressLink{http://www.sueddeutsche.de/wissen/psychologie-unterbewusstsein-durchschaut-unehrlichkeit-1.1923587}
The Times, Mar 26 \pressLink{http://www.thetimes.co.uk/tto/science/article4045032.ece}
Pacific Standard, Mar 25 \pressLink{http://www.psmag.com/navigation/health-and-behavior/unconscious-mind-better-detecting-lies-77368}
Science Daily, Mar 24 \pressLink{http://www.scidai.ly/releases/2014/03/140324104520.htm}
New Scientist, Mar 23 \pressLink{http://www.newscientist.com/article/mg22129610.700-invisible-how-to-see-through-lies.html}
} %
with potentially far--reaching practical
consequences. For example, consider jurors at court were advised
``Truth or lie --- trust your instinct, says research''
\cite{press_BBC_20140329,press_science_magazine_20140325}. This could
make it very difficult to allow for a rational debate in cases where
the truth does not seem as obvious as our instinct might suggest 
\cite{Loftus_03}. We show that the lie detection study does not
provide ``strong evidence'' that ``consciousness interfere[s] with the
natural ability to detect deception'' \cite[p.~6]{tenBrinke_14}.

The reasoning used by the lie detection study, as well as by many
other studies in the Neurosciences, is illustrated in
Fig.~1. Participants watched two videos of interrogations. In one
video the suspect was lying, in the other the suspect was telling the
truth.  Participants did not know who was the liar. The goal of the
study was to find out whether participants could tell apart liars from
truth--tellers (e.g., from signs of stress).  After watching the
videos, participants performed two tasks. The ``direct'' task
(Fig.~1A) is assumed to tap conscious processes because the
participants simply see pictures of the suspects and classify these
pictures as truth--tellers or liars. As expected
\cite{Bond_DePaulo_06}, participants were very bad in this direct task
(49.6\% correct, with chance level being 50\%).
\begin{figure*}[!tb]
  \centerline{\includegraphics{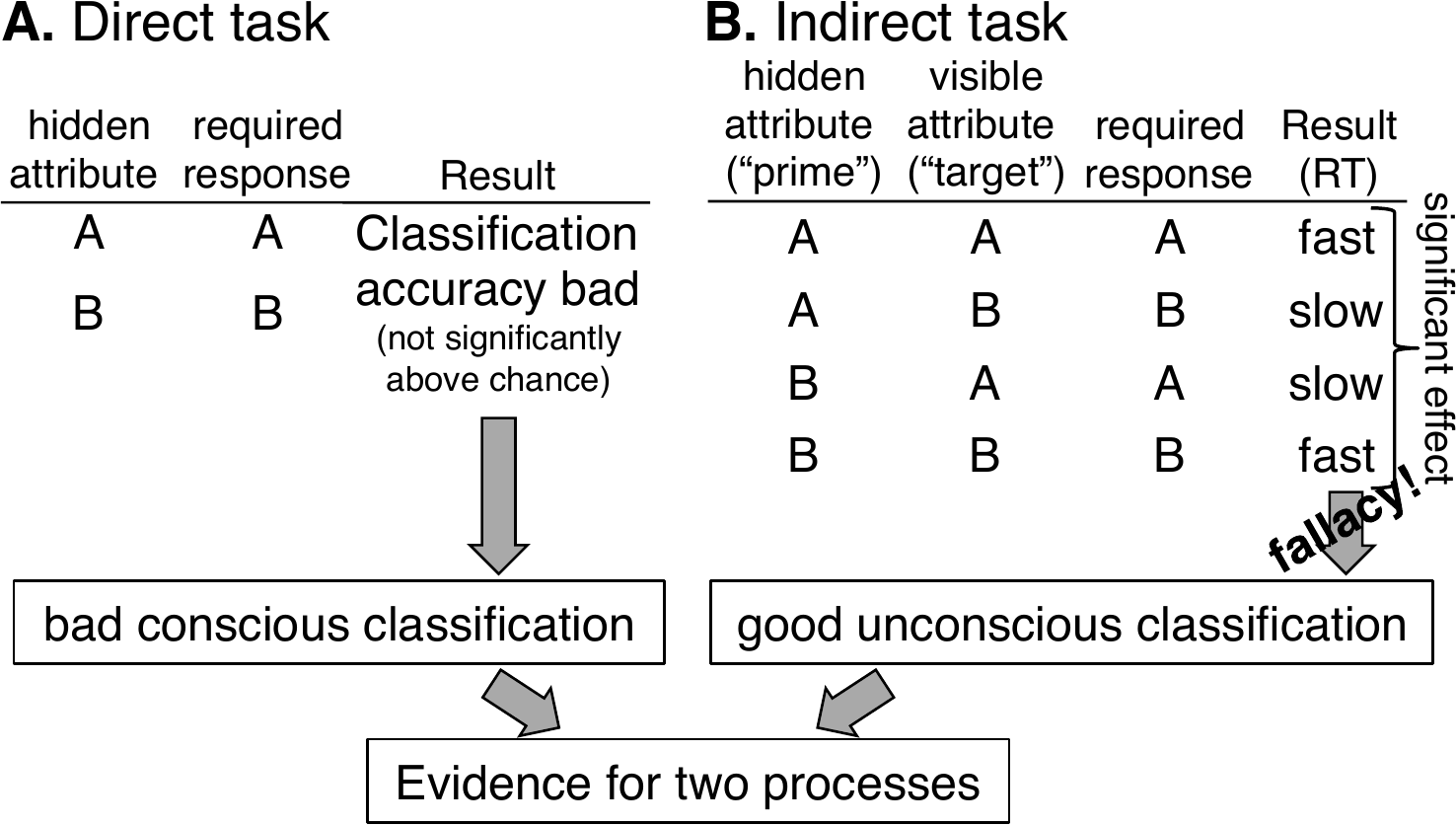}}
  \caption{Experimental rationale and fallacy: Typically there exists
    some hidden stimulus attribute. In the lie detection study this
    was whether the picture of a suspect showed a truth--teller or a
    liar. In other studies this could be the numerical size of a
    number or the emotional expression of a face that is hidden from
    consciousness by masking techniques. \textbf{A.}  Direct task:
    When participants directly classify the hidden attribute, they
    typically perform badly.  \textbf{B.}  Indirect task:
    Nevertheless, the hidden attribute (``prime'') can affect RTs if participants perform a task on another well visible
    stimulus (``target''). In the lie detection study, participants
    decided whether well visible target--words were related to lying
    or truth--telling. They were faster if the targets were preceded
    by a congruent but hidden picture (e.g., the word ``deceitful''
    preceded by the picture of a liar). While this is only possible if
    the hidden attribute was somehow processed by the nervous system,
    the fallacy is to conclude that there was relatively good
    unconscious classification accuracy of the hidden attribute,
    better than in the direct task.}
  \label{FigReasoning}
\end{figure*}

Things seemed to change drastically when the ``indirect'' task
(Fig.~1B) was performed, which is assumed to tap unconscious processes.
Now the pictures of the suspects were presented only briefly (``prime'')
and hidden
from consciousness by special masking techniques. The participants
sorted 
well visible words (the ``targets'') like ``honest'' or ``deceitful'' into
the categories ``truth'' or ``lie''. Interestingly, participants were
significantly faster if such a word was preceded by a congruent
picture of a suspect (e.g., the word ``deceitful'' was preceded by a
picture of a liar) than if the word was preceded by an incongruent
picture (e.g., the word ``deceitful'' was preceded by a picture of
truth--teller). This can only be explained if some information about
whether the masked picture shows a liar or a truth--teller has been
processed.

However, the authors of the lie detection study \cite{tenBrinke_14}
derived further reaching conclusions from the significant congruency
effect --- as is common practice in the Neurosciences.  They
concluded, that (i) the significant congruency effect indicates
``accurate unconscious assessments'' (p. 7) of truth--tellers vs.\
liars; %
(ii) in parallel to this accurate unconscious processing, there exists
another, inaccurate conscious process; %
(iii) the accurate unconscious assessments can even be ``made inaccurate
[...]  by conscious'' processes (p. 7), such that it might be wise to
prevent ``conscious deliberation about credibility'' (p. 7). 

We show
below that all these conclusions are not warranted by
the data. More generally, we describe that a significant congruency effect alone
does not provide sufficient evidence for such conclusions. 

\subsection*{The fallacy}
The main reason is that while the significant congruency effect indeed suggests that the
primes have been classified to a certain extent, it does not indicate
how good this classification was. The test for a significant
difference between reaction times (RTs) in congruent and incongruent trials is only
concerned with the question whether a `true' difference exists in the
population at all.  The test does not tell us how big this difference
is and for how good a classification performance it could be
harnessed. 

\paragraph*{In a nutshell:} 
The fallacy is to conclude from a significant effect in the
indirect task that there has been good indirect classification
performance of the prime (at least better than the classification
performance in the direct task). However, the significant effect only
indicates that {\em some} information about the stimuli has been
processed, not {\em how much} information. Given enough statistical
power, the indirect classification performance could be arbitrarily
small while nevertheless there could be a significant congruency
effect. This is not only a remote theoretical danger, as we show with
our reanalysis of the lie detection study. %

\paragraph*{Reanalysis of lie detection data} 

For the reanalysis, we put the data of the lie detection study to the
test: If the significant congruency effect on RTs is supposed to serve
as evidence for good unconscious processing, then we should be able to
use the RTs to decide for each trial whether the prime and target
stimuli were congruent or incongruent.  Small RTs would indicate a
congruent trial, large RTs would indicate an incongruent trial.

We applied  %
two classifiers to the data\footnote{Of the two experiments in the lie detection study
  \cite{tenBrinke_14}, we concentrate on the second one, as this is
  the one that presents `unconscious' stimuli. For the first
  experiment we obtained similar results (classification accuracy:
  51.1\%). All analyses were implemented twice independently, once in
  Matlab and once in R. The R--code is open available, see Part 1 of Materials
  and Methods.}: (i) the statistically optimal classifier
under the assumption that RTs follow normal or lognormal distributions
\cite{Ulrich_Miller_93} and (ii) a model--free classifier trained on
the data according to the standard protocol from statistical
learning (for details please consult the methods section). The two classifiers achieve classification accuracies of %
(i) 50.6\% and (ii) 49.3\%. %
We also found that (iii) on the given data there cannot exist a
classifier with accuracy larger than 54\% --- the same value that was
interpreted as ``detection incompetence'' in the lie detection study
\cite[p.~1]{tenBrinke_14}.  In short: the classification accuracy in
the unconscious task is just as dismal as in the conscious task and
can for all practical purposes be considered as being at chance level.
There is no evidence for ``accurate unconscious assessments''
\cite[p.~7]{tenBrinke_14}.

\begin{figure*}[!tb]
  \centerline{\includegraphics{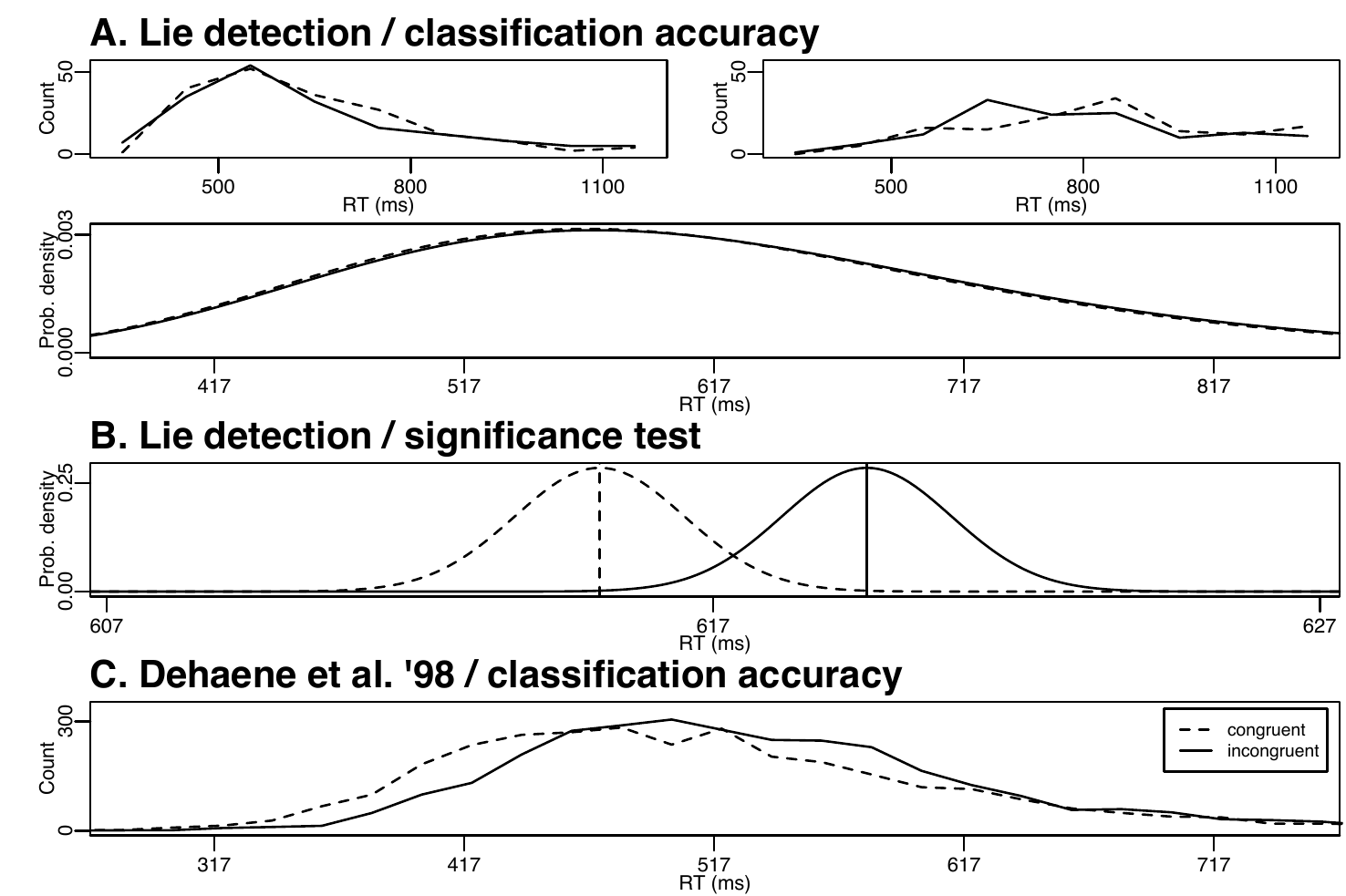}}
  \caption{RT-Distributions relevant for classification and
    significance tests. %
    \textbf{A.} Accurate classification of congruent vs.\ incongruent
    trials requires distinct RT--distributions. The top panels show
    RT-histograms for exemplary participants (left/right: participant
    with median/maximal accuracy of 50.8\%/56.1\%). %
    The large panel shows RT--distributions for an idealized
    participant, based on average values and lognormal distributions
    \protect\cite{Ulrich_Miller_93}. %
    All distributions overlap so heavily that classification accuracy
    is essentially at chance, showing that the RTs convey hardly any
    information about congruent vs.\ incongruent trials.  %
    \textbf{B.} A significant difference requires distinct
    distributions for the mean RTs of congruent vs.\ incongruent
    trials; with the standard deviation given by the standard error of
    the mean (SEM; e.g., \protect\citeNP{Franz_Loftus_12}).  These
    distributions are clearly distinct, reflecting the significant
    difference (\STt{65}{2.22}{p=0.03}; mean difference: 4.4~ms, SEM:
    2.0~ms, Cohen's d: 0.27; cf. \protect\citeNP{Cohen_88}). %
    Comparing A. and B. shows that classification accuracy can be at
    chance level even though the means are significantly
    different. This is caused by the massive reduction of the relevant
    standard deviation when calculating the SEM (cf. Part 3 in
    Materials \& Methods). Note, that we even had to change the scale
    of the abscissa in B to show the distributions appropriately.
    \textbf{C.} Histograms of RTs in behavioral task of
    \protect\citeA{Dehaene_etal_98}.  The distributions overlap
    heavily, suggesting that classification accuracy will be low. The
    histogram corresponds to Figure~2b of
    \protect\citeA{Dehaene_etal_98} and was electronically digitized
    from the printed version. In all plots, dashed/solid lines
    indicate congruent/incongruent conditions.}
  \label{FigData}
\end{figure*}

Fig.~2A illustrates this with the distributions relevant for
classification performance. The average RT--difference between
congruent and incongruent conditions was only
4.4~ms, 
whereas the average within--subjects standard deviation was 
146.5~ms. 
This gives a signal--to--noise ratio of 
0.03, 
which is much too small for a meaningful classification performance.

To understand why this can happen even though the RT means are
significantly different, note that the classification whether a trial
is congruent or incongrunent has to be performed on a single-trial
basis. In particular, the accuracy of the classifier does not improve
with more data. The statistical test for the difference in population
means, on the other hand, is based on the estimated variability of the
sample means, which gets smaller with more data. As shown in Fig.~2B,
it can easily happen that two distributions are nearly
indistinguishable by a classification task, yet a tiny difference in
their means becomes significant if the sample size or the number of
repetitions are large enough. See Part 2 in Materials \& Methods for more
details. 

\paragraph*{Better approaches.} 

What would a more appropriate approach look like? For a meaningful
comparison, we have to look not only for a significant effect, but
also at how much information is transmitted by this effect to the task
of classification.  A straightforward way to do this is to consider
the classification accuracy directly, as we did above. Other
approaches are possible as well. For example, one could use signal
detection theory \cite{Swets_61} on both tasks to determine and
compare appropriate d--prime values --- as has been done in some
studies \cite{Schmidt_02,Gegenfurtner_Franz_07,Schmidt_Vorberg_06}.
Alternatively, one could apply classic information theory on both
measures \cite{Shannon_1948}, an approach we are currently working on.
For the lie detection study, all these methods would lead to the same
conclusion: unconscious lie detection does not work any better than
its conscious counterpart. Both are essentially at chance--level. 

\subsection*{The problematic reasoning is widely used}

One might argue that this is a limited problem of one single
study. However, the problematic reasoning is widely and routinely
used. For illustration, we sketch three highly influential
studies \cite{Dehaene_etal_98,Morris_etal_98,Pessiglione_etal_07}.
Many more studies exist in the literature. 

\citeA{Dehaene_etal_98} investigated whether humans can unconsciously
process information about the magnitude of numbers. Stimuli were
numbers between 1 and 9 that were hidden from consciousness by
masking. Participants categorized whether the numbers were larger or
smaller than 5. In direct tasks (Fig.~1A) participants were not
significantly different from chance level (52.6\% and 54\%
correct). %
Nevertheless, the masked numbers had significant effects in indirect
tasks (Fig.~1B): If participants responded to a target number that
could be congruent with the prime (e.g., both smaller than 5) or
incongruent (e.g., one smaller and the other larger than 5), then
participants showed significant effects on RTs and significant
lateralizations in electroencephalography (EEG) and functional
magnetic resonance imaging (fMRI). Based on the same reasoning as
outlined above, \citeA{Dehaene_etal_98} concluded that in
the indirect task participants ``unconsciously appl[ied] the task
instructions to the prime, would therefore categorize it as smaller or
larger than 5, and would even prepare a motor response appropriate to
the prime'' (p.  598). The authors summarize ``that a large amount of
cerebral processing [...]  can be performed in the absence of
consciousness'' (p.  599).

However, these significant differences in RTs, EEG and fMRI
measurements do not tell whether classification accuracy in the
indirect task was better than in the direct task. If not, then there
would be no evidence for unconscious processing of the primes. We
cannot directly evaluate the relevant classification performance
because we do not have access to the data. Instead, we analyzed the
published histogram of all RTs performed in the behavioral task
(Fig.~2C). If we determine the classification accuracy based on these
distributions, we obtain 55\% correct,
which is discomfortingly close to the accuracy in the direct tasks.
While this is only a very rough estimate, it cannot rule out the
possibility that there might indeed be a similar problem in the study
by \citeA{Dehaene_etal_98} as we found for the lie
detection study.  The only way to find out would be replication
studies or a reanalysis of the existing data.

\citeA{Morris_etal_98} investigated emotional learning in
the amygdala. Two angry faces were used as stimuli, one of which had
been conditioned to an aversive event. The faces were hidden from
consciousness by masking, such that participants were at chance when
classifying whether such a face was shown to them.  Nevertheless,
activity in the right amygdala was significantly modulated by the fact
that one of the two faces had been associated with the aversive event,
as measured with positron emission tomography (PET).  Using again the
same reasoning, the authors conclude that ``we provide the first
evidence that the human amygdala can discriminate the acquired
behavioral significance of stimuli without the need for conscious
perception'' (p. 469). Our critique is again: The significant
modulation of amygdala activity does not show whether there is also
good classification accuracy that is clearly different from chance
level and that would justify the conclusion of a superior process
operating in parallel to the conscious process.

\citeA{Pessiglione_etal_07} investigated subliminal
motivation. Images of coins were presented, either one pound or one
penny, and hidden from consciousness by masking, such that
participants were at chance level when classifying the coins.
Nevertheless, activity in the ventral pallidum (VP) was significantly
modulated by the value of the coins, as measured by fMRI (similar
results were found for skin conductance and grip force).  The authors
concluded that there are two motivational processes, one conscious and
the other unconscious: ``Thus, only the VP appeared in position to
modulate behavioral activation according to subliminal incentives and
hence to underpin a low--level motivational process, as opposed to a
conscious cost--benefit calculation'' (p.  906). Our concerns are
again the same: The significant modulation of activation does not tell
whether the information available to the VP suffices for a
classification performance that is clearly better than the conscious
classification performance. Therefore, it is not clear whether the
authors' assumption of two processes (an unconscious and a conscious
one) for cost--benefit calculation is warranted.

\subsection*{Is there unconscious processing?} 

Because all our example studies 
happen to be related to the question of whether there exists
unconscious processing independent of and parallel to conscious
processing, we want to preclude a potential misunderstanding.  We are
mainly interested in describing the methodological fallacy, not in
discussing unconscious processing. Such a discussion would go beyond
the scope of this article and would have to take into account a long
history of research
\cite{Eriksen_60,Holender_86,Reingold_Merikle_88,Greenwald_etal_96,Hannula_etal_05,Kouider_Dehaene_07}.
Therefore, we do not claim that unconscious processing independent of
conscious processing does not exist or cannot be shown. We do,
however, claim that the lie detection study does not provide evidence
for a superior unconscious lie detection ability\footnote{Note that
  our conclusions on the lie detection study \cite{tenBrinke_14} are
  corroborated by a recent commentary of lie--detection experts
  \cite{Levine_Bond_14} who question the plausibility of the
  lie-detection results in the light of other research and
  meta--analyses in this area. While \citeA{Levine_Bond_14} had to
  speculate that one of the conditions is a statistical outlier we can
  now show the statistical reasons behind the wrong
  conclusions. Hence, our findings converge with the intuition and
  meta--analytic data of these lie--detection experts.} %
and that this study shows in an exemplary way how the claims of the
other studies using the same flawed rationale can go astray and need
careful reconsideration using more appropriate methods

\subsection*{Conclusions} 

We described a reasoning that is widely used but flawed. In the case
of the lie detection study \cite{tenBrinke_14}, the commendable
open--data practice allowed us to show in an exemplary way how this
reasoning can lead to wrong conclusions. More generally, conclusions
of the many studies using this reasoning should be treated with
caution and could be wrong.  In the future, we should employ better
statistical methods and conclusions based on the flawed reasoning
should be reconsidered.

\newpage 
\renewcommand{\thesection}{}
\renewcommand{\thesubsection}{\arabic{subsection}}
\section*{Materials and Methods}

\subsection{Classification and statistical optimality in more detail}
\label{OptimalClassification}

In this section, we describe how an optimal classification of the
single trial data in an indirect task (Fig.~1B of main text) can be
performed.  We prove below that under the assumption that the RTs
follow a normal distribution or a lognormal
distribution  \cite{Ulrich_Miller_93}, the statistically optimal
classifier is given by a median split of the reaction times (``median
classifier'').  We also describe a typical classifier as used in
machine learning that does not require any distributional assumptions
(``trained classifier''). Finally, we derive a theoretical upper bound
for classification performance on the given data that in principle can
never be exceeded (``over--optimistic upper bound'').  Before going
into details, we first describe the results of applying these
classifiers to the data of the lie detection study  \cite{tenBrinke_14}.

\paragraph*{Classification results for lie detection study.} For each
participant, the goal is to classify the trials in the indirect task
as 'congruent' or 'incongruent', based on the RTs of this participant.
For each participant, we proceed as follows.
(i) For the model--based median classifier, we compute the median RT,
use this as the threshold of a step function classifier (see below),
and compute the accuracy of this classifier over all trials.
(ii) For the model--free trained classifier, we randomly split the
trials into a training and test set of 50\% each (other split sizes
lead to very similar results). We determine the best threshold on the
training set, and compute the resulting accuracy on the test set. We
repeat this procedure 10 times with different random splits of the
data and report the average over these test accuracies.
(iii) For the over--optimistic upper bound, we evaluate the accuracy of
all possible thresholds for the step function classifier over all
trials and report the best result.
The following table shows means and standard deviations over the
accuracies of all participants:

\bigskip
\noindent
\begin{tabular}{lccl}
\bf Method                             & \bf mean(accuracy) & \bf std(accuracy) \\
\hline
(i) Median classifier (model: lognormal)          & 50.61\%  &  2.65\%\\
 \hspace{0.6cm}Median classifier (model: normal) & 50.61\%  &  2.65\%\\
 (ii) Trained classifier (model--free)            & 49.34\%  &  2.64\% \\
\hline
(iii) Over--optimistic  upper bound               & 53.73\%  &  1.99\%\\
\end{tabular}

\bigskip
\noindent We can see that both the model--based (i) and model--free
(ii) classifiers perform nearly exactly at chance level. The
over--optimistic upper bound shows that on this data set, there does
not exist a classifier that can obtain an accuracy higher than 54\%
--- the value that was interpreted as ``detection incompetence'' in
the lie detection study \cite[p.~1]{tenBrinke_14}.

\paragraph*{General form of the optimal classifier.}  Consider a
classification task where the input is a real-valued number $x$ (e.g.,
a reaction time, RT), and the classifier is supposed to predict one of
two labels $y$ (e.g., 'congruent' or 'incongruent'; for simplicity we
use labels 1 and 2 in the following). Following the standard setup in
statistical decision theory \cite[section 1.5]{Bishop06} we assume
that the input data $X$ and the output data $Y$ are drawn according to
some fixed (but unknown) probability distribution $P$.  This
distribution can be described uniquely by the class-conditional
distributions $P( X \condon Y = 1)$ and $P(X \condon Y = 2)$ and the
class priors $\pi_1 = P(Y = 1)$ and $\pi_2 = P(Y=2)$. A classifier is
a function $f:\R \to \{1,2\}$ that assigns a label $y$ to each input
$x$. The classifier that has the smallest probability of error is
called the Bayes classifier. In case the classes have equal weight,
that is $\pi_1 = \pi_2$, the Bayes classifier has a particularly
simple form: it classifies an input point $x$ by the class that has
the higher class-conditional density at this point. Formally, this
classifier is given by

\banum
\label{eq-fopt}
f_{opt}(x) := \begin{cases} 1 & \text{ if } P(X = x \condon Y =1 ) > P(X = x \condon Y=2)\\
2 & \text{ otherwise.} 
\end{cases}
\eanum

\paragraph*{Optimal classifier for normal and lognormal
  distributions.}  We now consider the special case where the class-conditionals follow a particular
distribution. Let us start with the normally distributed case. We assume
that both class-conditionals are normal distributions with means
$\mu_1$, $\mu_2$ and equal variance $\sigma^2$, 
and we denote their corresponding probability
density functions (pdfs) by $\varphi_{\mu_1,\sigma}$ and
$\varphi_{\mu_2,\sigma}$. Under the additional assumption that  both classes
have equal weights $\pi_1 = \pi_2 = 0.5$, the 
cumulative distribution 
function (cdf) of the input (marginal distribution of $X$)  is given as 
\banum
\label{eq-gaussian}
&\Gamma(x) := 0.5 \cdot \Big( \Phi(\frac{x - \mu_1}{\sigma}) +
\Phi(\frac{x - \mu_2}{\sigma} )\Big), 
\eanum
where $\Phi$ denotes the cdf of the standard normal distribution. For
$t \in \R$, we introduce the step function classifier with threshold
$t$ by 
\banum \label{eq-step}
f_t(x) := \begin{cases} 1 & \text{ if } x \leq t\\
2 & \text{ otherwise.} 
\end{cases}
\eanum
In the special case where the threshold $t$ coincides with the median of
the marginal distribution of $X$, we call the resulting step function
classifier the {\em median classifier. }

\begin{propositionnn}[Median classifier is optimal for normal model]
  If the input distribution is given by Eq.~\eqref{eq-gaussian}, then 
the optimal classifier $f_{opt}$ coincides with the median
classifier. 
\end{propositionnn}
{\em Proof.}
Because both classes have the same weight of 0.5, the
Bayes classifier is given by $f_{opt}$ as in Eq.~\eqref{eq-fopt}. 
For any choice of $\mu_1$, $\mu_2$ and $\sigma$, the class-conditional
pdfs $\varphi_{\mu_1, \sigma}$ and $\varphi_{\mu_2,\sigma}$  intersect
exactly once, namely at $t^* = (\mu_1 + \mu_2) / 2$. By definition of
$f_{opt}$, the optimal classifier $f_{opt}$ is then the step function
classifier with threshold $t^*$. We now compute the value of the cdf at 
$t^*$: 

\ba
\Gamma(t^*) & = 0.5 \cdot 
\Big( \Phi(\frac{t^* - \mu_1}{\sigma}) + \Phi(\frac{t^* - \mu_2}{\sigma} )\Big)\\
&= 0.5 \cdot  
\Big( \Phi(\frac{\mu_2 - \mu_1}{2\sigma}) + \Phi(\frac{\mu_1 - \mu_2}{2\sigma})\Big)\\
&= 0.5 \cdot  
\Big(\Phi(\frac{\mu_2 - \mu_1}{2}) + (1 - \Phi(\frac{\mu_2 - \mu_1}{2})\Big)\\
& = 0.5. 
\ea
Here, the second last equality comes from the fact that the normal
distribution is symmetric about 0.  This calculation shows that the
optimal threshold $t^*$ indeed coincides with the median of the input
distribution, which is what we wanted to prove. \hfill$\Box$

It is easy to see that this proof can be generalized to more general
types of symmetric probability distributions. It is, however, even
possible to prove an analogous statement for lognormal distributions,
which are not symmetric themselves. We introduce the notation
$\lambda_{\mu,\sigma}$ for the probability density function (pdf) of a
lognormal distribution, and $\Lambda_{\mu,\sigma}$ for the
corresponding cdf. These functions are defined as
\ba
&\lambda_{\mu,\sigma}(x) :=  
\frac{1}{x \sigma \sqrt{2\pi}}
\exp\Big(- \frac{(\log x  -\mu)^2 }{2 \sigma^2} \Big) 
&& \text{ and } &&\Lambda_{\mu,\sigma}(x) := \Phi\Big(     
\frac{\log x  -\mu}{\sigma}
\Big).
\ea

Consider the case where the class-conditional distributions
are lognormal distributions with same scale parameter $\sigma$ but
different location parameters $\mu_1$ and $\mu_2$, and 
assume that both classes have the same weights $\pi_1 = \pi_2 =
0.5$. Then the pdf and cdf of the input distribution (marginal
distribution of $X$) are given as
\banum
& g(x) = 0.5 \cdot \;(\; \lambda_{\mu_1, \sigma}(x) + \lambda_{\mu_2,
  \sigma}(x) \;) \nonumber \\
& G(x) = 0.5 \cdot \;(\; \Lambda_{\mu_1, \sigma}(x) + \Lambda_{\mu_2,
  \sigma}(x) \;). \label{eq-lognormal}
\eanum
\begin{propositionnn}[Median classifier is optimal for lognormal
  model] 
  If the input distribution is given by Eq.~\eqref{eq-lognormal}, then 
the optimal classifier $f_{opt}$ coincides with the median
classifier. 
\end{propositionnn}
{\em Proof.} %
The proof is analogous to the previous one. 
For any choice of $\mu_1$, $\mu_2$ and $\sigma$, the densities
$\lambda_{\mu_1, \sigma}$ and $\lambda_{\mu_2, \sigma}$ intersect
exactly once. To see this, we solve the equation $\lambda_{\mu_1,
  \sigma}(t^*) = \lambda_{\mu_2, \sigma}(t^*)$, which leads to the
unique solution $t^*= \exp( (\mu_1 + \mu_2)
/ 2)$. 
The input cdf at this value can be computed as 
\ba
G(t^*) & = 0.5 \Big( \Lambda_{\mu_1,
  \sigma}(t^*) +  \Lambda_{\mu_2, \sigma}(t^*) \Big) \\
& =  0.5\Big( \Phi( \frac{\mu_2 - \mu_1}{2\sigma} )+  \Phi( \frac{\mu_1 - \mu_2}{2\sigma} ) \Big)\\
& = 0.5. 
\ea
The last step follows as above by the symmetry of the normal cdf. 
\hfill$\Box$

\paragraph*{Training a model--free classifier. } If we do not want to
make any assumptions about the underlying probability distribution, we can
follow the standard protocol of statistical learning to identify the
threshold $t$ of the best step function classifier. For each
participant, we are given trials in form of input-output pairs
$(X_i, Y_i)_{i=1,...,n}$, $X_i \in \R$, $Y_i \in \{1, 2\}$. We
randomly split this data set into a training set consisting of 50\%
and a test set of the remaining 50\% of all trials. On the training
set, we determine the threshold $t^*$ that leads to the smallest
number of misclassifications (= training error). For the corresponding
step function classifier $f_{t^*}$ we now compute the error on the
test set (= test error). We repeat this procedure 10 times to remove potential
subsampling artifacts and report the mean over these repetitions. 
For readers familiar with machine learning,
note that in this simple scenario, no model selection is involved, so a
more complex evaluation procedure 
such as cross validation is not necessary.

\paragraph*{An over--optimistic upper bound on classification accuracy.} 
To rule out the case that the result of the model--free classifier is
seriously sub-optimal (due to the effect of splitting the data in
training and test sets, or due to overfitting or underfitting), we can
derive an upper bound on the accuracy of the best step function classifier that
possibly exists on the given data. For each participant, we cycle
through all possible thresholds $t$ and evaluate the accuracy of the
corresponding step function classifier $f_t$ on all trials. We then
select the best accuracy obtained in this way as the classification
accuracy of this participant. This accuracy is overly optimistic, as
this classifier usually overfits and exploits sampling artifacts. On
the other hand, it gives an upper bound on the classification accuracy
that any other step function classifier could potentially achieve on
the data. Finally, note that in the context of the RT experiment, 
it would not make sense to consider classifiers that do not have the
form of a step function classifier --- the general classification
scenario implied by the experimental setup is to
separate slow RTs from fast RTs. 

\subsection{Why is the relevant standard deviation for the
  significance test much smaller than that for classification?}
\label{IntuitiveExplanation}

Let us illustrate our answer with the data of the lie detection
study \cite{tenBrinke_14}. Consider two probability distributions with
slightly different expected values, such as the ones in Fig.~2A.
The task of the classifier is to predict for each trial whether the
measured RT has been generated from a congruent or an incongruent
condition. More abstractly, given a real-valued sample, we want to
decide which of the two distributions is more likely to have generated
that sample point. In general, this will only be possible in a
satisfactory manner if the two distributions have only little overlap
and their means are considerably different from each other.

The significance test, on the other hand, assesses whether the
expected values of the two distributions are different at all. It does
not ask for a large difference, it just asks for any difference. The
more measurements are taken, the closer each mean estimate will be to
the corresponding expected value. We know from the central limit
theorem that the SEM is of order $1 / \sqrt{n}$. In the limiting case
of an infinite number of measurements, the SEM would approach zero.

To get a feeling for this effect, consider the data of the lie
detection study. For a rough estimate, let us for a moment ignore the
fact that there were different participants (i.e., that
between--subjects variability exists; this is not so critical because
we are dealing with within--subjects designs such that the difference
mainly is affected by within--subjects variability
\cite{Franz_Loftus_12}).  The average standard deviation for a trial
was 146.5~ms (Fig.~2A). Each condition was measured about 180 times
in each participant and the study had 66~participants, which leads to
a factor of ${1}/{\sqrt{180 \cdot 66}}$. Taking the difference between
the congruent and incongruent conditions increases the SEM by a factor
of $\sqrt{2}$ (assuming for simplicity independence and equal
variances), such that a rough prediction for the SEM relevant for the
significance test is given as $146.5\cdot \sqrt{2} / \sqrt{180 \cdot
  66 }\mbox{ ms}= 1.9$~ms. This is close to the empirically obtained
$2.0$~ms.

\subsection{If the means are significantly different, doesn't this
  imply that the classification accuracy is significantly different
  from chance level? }
\label{ss-classification-significant}

If we have enough statistical power, significance tests will
eventually show that classification is different from chance if the
means are significantly different. (In real data, both significances
might not occur at the same time because sources of noise are not exactly identical for the
means and the classification results.) %

However, with regard to the reasoning outlined in Fig.~1 of the main
paper, this question is misleading. For the typical neuroscientific
interpretation it is not only important that the classification
accuracy in the indirect task is significantly different from chance
level (this could also happen if the true classification performance
were, say, 51\%). What counts is whether the classification accuracy
is {\em considerably larger} than chance level, and in particular,
considerably larger than the accuracy obtained in the direct task.
This leads back to the old statistical issue of needing to distinguish
between the statistical significance of effects vs.\ the size of the
effects.

\subsection{Many studies did not test for the difference of the
  effects.  Isn't this also a problem?}
\label{TestForDifference}

The correct procedure would indeed be to test for the difference
\cite{Franz_Gegenfurtner_08,Nieuwenhuis_etal_11}. This is, however, an
issue independent of the general fallacy we are concerned with, so we
do not discuss it further.

\subsection{The lie detection study calculated Cohen's d values.
  Doesn't this ameliorate the problem?}
\label{CohenD}

While most studies used the RTs in the indirect task for their
significance test, the lie detection study \cite{tenBrinke_14} used a
somewhat different approach. For each participant an individual
Cohen's d value \cite{Cohen_88} was computed from the RTs, resulting
for the $66$ participants in $d_1, ..., d_{66}$. Then, a significance
test was performed on these values and a second Cohen's d value
$d_{across}$ was calculated across the individual values $d_1, ...,
d_{66}$.  This value $d_{across}$ is what is shown in Fig.~2
(Exp.~2) of the lie detection study \cite{tenBrinke_14}, it was
found to be $d_{across} = 0.27$ \cite[p.~6]{tenBrinke_14}. %
This means that the Cohen's d--values used in this figure refer to the
question of whether the means of the RTs are different from each other
(which they very well can be, as we explained above). It does not say
anything about the effect size of the indirect classification
performance, which would be the relevant quantity. The relevant
Cohen's d--values computed on the distribution that is relevant to
classification (our Fig.~2A) amount on average to $0.03$
\cite[p.~6]{tenBrinke_14}, %
which is very small and therefore fully consistent with the results of
our classifiers (for pychophysicists: this is equivalent to a very
small d--prime value in signal detection theory).
\newpage
\section*{Acknowledgments}
We thank the authors \cite{tenBrinke_14} and the editor \cite{Eich_14}
of the lie detection study for their open data policy and regret that
due to this commendable openness they are the first to be criticized
for a method that has been used in many studies.  We thank Gilles
Blanchard and Frank R{\"o}sler for comments on the manuscript.
U.v.L. was supported by the German Research Foundation (grant
LU1718/1-1 and Research Unit 1735 ``Structural Inference in
Statistics: Adaptation and Efficiency'').

\subsection*{Author contributions}
The major contribution to this manuscript comes from V.H.F. He
discovered the methodological flaws and reanalyzed the data of the lie
detection study \cite{tenBrinke_14}. The role of U.v.L. was the one of
a critical discussion partner. She verified all arguments and
re-implemented the analyses independently in Matlab. Both authors
jointly wrote the paper.

\subsection*{Competing Interests}
Volker H. Franz received funding from the University of Hamburg and
the German Research Foundation. He currently serves in the editorial
board of British Journal of Psychology. Ulrike von Luxburg received
funding from the University of Hamburg and the German Research
Foundation. She currently serves in the editorial board of the Journal
of Machine Learning Research and is a board member of the
International Machine Learning Society. Previously she served in the
editorial board of Statistics and Computing.

\newpage

\begin{thebibliography}{}

\bibitem[\protect\BCAY{Anonymous}{Anonymous}{2014}]{press_science_magazine_20140325}
Anonymous.
\newblock{}\BBOP{}2014\BBCP{}.
\newblock{}\Bem{{To spot a liar, trust your instinct}.}
\newblock{}Science Magazine (online: March 25, 2014) {\tt
  http://news.sciencemag.
  org/signal-noise/2014/03/spot-liar-trust-your-instinct}.

\bibitem[\protect\BCAY{Bishop}{Bishop}{2006}]{Bishop06}
Bishop, C.
\newblock{}\BBOP{}2006\BBCP{}.
\newblock{}\Bem{Pattern recognition and machine learning.}
\newblock{}Springer.

\bibitem[\protect\BCAY{Bond \BBA{} DePaulo}{Bond \BBA{}
  DePaulo}{2006}]{Bond_DePaulo_06}
Bond, C.~F.\BCBT{} \BBA{} DePaulo, B.~M.
\newblock{}\BBOP{}2006\BBCP{}.
\newblock{}\BBOQ{}{Accuracy of deception judgments}.\BBCQ{}
\newblock{}\Bem{Personality and Social Psychology Review}, \Bem{10}, 214--234.

\bibitem[\protect\BCAY{Briggs}{Briggs}{2014}]{press_BBC_20140329}
Briggs, H.
\newblock{}\BBOP{}2014\BBCP{}.
\newblock{}\Bem{{Truth or lie --- trust your instinct, says research}.}
\newblock{}British Broadcasting Corporation (online: March 29, 2014, {\tt
  http://www.bbc.com/news/health-26764866}).

\bibitem[\protect\BCAY{Cohen}{Cohen}{1988}]{Cohen_88}
Cohen, J.
\newblock{}\BBOP{}1988\BBCP{}.
\newblock{}\Bem{{Statistical power analysis for the behavioral sciences}}\
  (2nd\ \BEd).
\newblock{}Hillsdale, NJ: Erlbaum.

\bibitem[\protect\BCAY{Dehaene \BOthers{}}{Dehaene
  \BOthers{}}{1998}]{Dehaene_etal_98}
Dehaene, S., Naccache, L., {Le Clec'H}, G., Koechlin, E., Mueller, M.,
  Dehaene-Lambertz, G., {van de Moortele}, P.~F.\BCBL{} \BBA{} {Le Bihan}, D.
\newblock{}\BBOP{}1998\BBCP{}.
\newblock{}\BBOQ{}{Imaging unconscious semantic priming}.\BBCQ{}
\newblock{}\Bem{Nature}, \Bem{395}, 597--600.

\bibitem[\protect\BCAY{Eich}{Eich}{2014}]{Eich_14}
Eich, E.
\newblock{}\BBOP{}2014\BBCP{}.
\newblock{}\BBOQ{}{Business not as usual}.\BBCQ{}
\newblock{}\Bem{Psychological Science}, \Bem{25}(1), 3--6.

\bibitem[\protect\BCAY{Eriksen}{Eriksen}{1960}]{Eriksen_60}
Eriksen, C.~W.
\newblock{}\BBOP{}1960\BBCP{}.
\newblock{}\BBOQ{}Discrimination and learning without awareness --- a
  methodological survey and evaluation.\BBCQ{}
\newblock{}\Bem{Psychological Review}, \Bem{67}, 279--300.

\bibitem[\protect\BCAY{Franz \BBA{} Gegenfurtner}{Franz \BBA{}
  Gegenfurtner}{2008}]{Franz_Gegenfurtner_08}
Franz, V.~H.\BCBT{} \BBA{} Gegenfurtner, K.~R.
\newblock{}\BBOP{}2008\BBCP{}.
\newblock{}\BBOQ{}{Grasping visual illusions: Consistent data and no
  dissociation}.\BBCQ{}
\newblock{}\Bem{Cognitive Neuropsychology}, \Bem{25}(7), 920--950.

\bibitem[\protect\BCAY{Franz \BBA{} Loftus}{Franz \BBA{}
  Loftus}{2012}]{Franz_Loftus_12}
Franz, V.~H.\BCBT{} \BBA{} Loftus, G.~R.
\newblock{}\BBOP{}2012\BBCP{}.
\newblock{}\BBOQ{}{Standard errors and confidence intervals in within--subjects
  designs: Generalizing Loftus \& Masson (1994) and avoiding biases of
  alternative accounts}.\BBCQ{}
\newblock{}\Bem{Psychonomic Bulletin \& Review}, \Bem{19}(3), 395-404.

\bibitem[\protect\BCAY{Gegenfurtner \BBA{} Franz}{Gegenfurtner \BBA{}
  Franz}{2007}]{Gegenfurtner_Franz_07}
Gegenfurtner, K.~R.\BCBT{} \BBA{} Franz, V.~H.
\newblock{}\BBOP{}2007\BBCP{}.
\newblock{}\BBOQ{}{A comparison of localization judgments and pointing
  precision}.\BBCQ{}
\newblock{}\Bem{Journal of Vision}, \Bem{7}, 1--12.

\bibitem[\protect\BCAY{Greenwald, Draine\BCBL{} \BBA{} Abrams}{Greenwald
  \BOthers{}}{1996}]{Greenwald_etal_96}
Greenwald, A.~G., Draine, S.~C.\BCBL{} \BBA{} Abrams, R.~L.
\newblock{}\BBOP{}1996\BBCP{}.
\newblock{}\BBOQ{}{Three cognitive markers of unconscious semantic
  activation}.\BBCQ{}
\newblock{}\Bem{Science}, \Bem{273}, 1699--1702.

\bibitem[\protect\BCAY{Hannula, Simons\BCBL{} \BBA{} Cohen}{Hannula
  \BOthers{}}{2005}]{Hannula_etal_05}
Hannula, D.~E., Simons, D.~J.\BCBL{} \BBA{} Cohen, N.~J.
\newblock{}\BBOP{}2005\BBCP{}.
\newblock{}\BBOQ{}Imaging implicit perception: promise and pitfalls.\BBCQ{}
\newblock{}\Bem{Nature Reviews Neuroscience}, \Bem{6}(3), 247--255.

\bibitem[\protect\BCAY{Holender}{Holender}{1986}]{Holender_86}
Holender, D.
\newblock{}\BBOP{}1986\BBCP{}.
\newblock{}\BBOQ{}Semantic activation without conscious identification in
  dichotic--listening, parafoveal vision, and visual masking --- a survey and
  appraisal.\BBCQ{}
\newblock{}\Bem{Behavioral and Brain Sciences}, \Bem{9}, 1-23.

\bibitem[\protect\BCAY{Kouider \BBA{} Dehaene}{Kouider \BBA{}
  Dehaene}{2007}]{Kouider_Dehaene_07}
Kouider, S.\BCBT{} \BBA{} Dehaene, S.
\newblock{}\BBOP{}2007\BBCP{}.
\newblock{}\BBOQ{}{Levels of processing during non--conscious perception: a
  critical review of visual masking}.\BBCQ{}
\newblock{}\Bem{Philosophical Transactions of the Royal Society B ---
  Biological Sciences}, \Bem{362}, 857--875.

\bibitem[\protect\BCAY{Levine \BBA{} Bond}{Levine \BBA{}
  Bond}{2014}]{Levine_Bond_14}
Levine, T.~R.\BCBT{} \BBA{} Bond, C.~F.
\newblock{}\BBOP{}2014\BBCP{}.
\newblock{}\BBOQ{}{Direct and indirect measures of lie detection tell the same
  story: A reply to ten Brinke, Stimson, and Carney (2014)}.\BBCQ{}
\newblock{}\Bem{Psychological Science}.
\newblock{}(published online June 25, 2014)

\bibitem[\protect\BCAY{Loftus}{Loftus}{2003}]{Loftus_03}
Loftus, E.
\newblock{}\BBOP{}2003\BBCP{}.
\newblock{}\BBOQ{}{Science and society --- Our changeable memories: Legal and
  practical implications}.\BBCQ{}
\newblock{}\Bem{Nature Reviews Neuroscience}, \Bem{4}, 231--234.

\bibitem[\protect\BCAY{Morris, {\"O}hman\BCBL{} \BBA{} Dolan}{Morris
  \BOthers{}}{1998}]{Morris_etal_98}
Morris, J.~S., {\"O}hman, A.\BCBL{} \BBA{} Dolan, R.~J.
\newblock{}\BBOP{}1998\BBCP{}.
\newblock{}\BBOQ{}{Conscious and unconscious emotional learning in the human
  amygdala}.\BBCQ{}
\newblock{}\Bem{Nature}, \Bem{393}, 467--470.

\bibitem[\protect\BCAY{Nieuwenhuis, Forstmann\BCBL{} \BBA{}
  Wagenmakers}{Nieuwenhuis \BOthers{}}{2011}]{Nieuwenhuis_etal_11}
Nieuwenhuis, S., Forstmann, B.~U.\BCBL{} \BBA{} Wagenmakers, E.~J.
\newblock{}\BBOP{}2011\BBCP{}.
\newblock{}\BBOQ{}{Erroneous analyses of interactions in neuroscience: A
  problem of significance}.\BBCQ{}
\newblock{}\Bem{Nature Neuroscience}, \Bem{14}, 1105--1107.

\bibitem[\protect\BCAY{Pessiglione \BOthers{}}{Pessiglione
  \BOthers{}}{2007}]{Pessiglione_etal_07}
Pessiglione, M., Schmidt, L., Draganski, B., Kalisch, R., Lau, H., Dolan,
  R.~J.\BCBL{} \BBA{} Frith, C.~D.
\newblock{}\BBOP{}2007\BBCP{}.
\newblock{}\BBOQ{}How the brain translates money into force: {A} neuroimaging
  study of subliminal motivation.\BBCQ{}
\newblock{}\Bem{Science}, \Bem{316}, 904--906.

\bibitem[\protect\BCAY{Reingold \BBA{} Merikle}{Reingold \BBA{}
  Merikle}{1988}]{Reingold_Merikle_88}
Reingold, E.~M.\BCBT{} \BBA{} Merikle, P.~M.
\newblock{}\BBOP{}1988\BBCP{}.
\newblock{}\BBOQ{}{Using direct and indirect measures to study perception
  without awareness}.\BBCQ{}
\newblock{}\Bem{Perception {\&} Psychophysics}, \Bem{44}(6), 563--575.

\bibitem[\protect\BCAY{Schmidt}{Schmidt}{2002}]{Schmidt_02}
Schmidt, T.
\newblock{}\BBOP{}2002\BBCP{}.
\newblock{}\BBOQ{}{The fingers in flight: Real--time control by visually masked
  color stimuli}.\BBCQ{}
\newblock{}\Bem{Psychological Science}, \Bem{13}(2), 112--118.

\bibitem[\protect\BCAY{Schmidt \BBA{} Vorberg}{Schmidt \BBA{}
  Vorberg}{2006}]{Schmidt_Vorberg_06}
Schmidt, T.\BCBT{} \BBA{} Vorberg, D.
\newblock{}\BBOP{}2006\BBCP{}.
\newblock{}\BBOQ{}Criteria for unconscious cognition: {Three} types of
  dissociation.\BBCQ{}
\newblock{}\Bem{Perception \& Psychophysics}, \Bem{68}, 489--504.

\bibitem[\protect\BCAY{Shannon}{Shannon}{1948}]{Shannon_1948}
Shannon, C.~E.
\newblock{}\BBOP{}1948\BBCP{}.
\newblock{}\BBOQ{}A mathematical theory of communication.\BBCQ{}
\newblock{}\Bem{The Bell System Technical Journal}, \Bem{27}, 379--423 and
  623--656.

\bibitem[\protect\BCAY{Swets}{Swets}{1961}]{Swets_61}
Swets, J.~A.
\newblock{}\BBOP{}1961\BBCP{}.
\newblock{}\BBOQ{}{Is there a sensory threshold?}\BBCQ{}
\newblock{}\Bem{Science}, \Bem{134}(3473), 168--177.

\bibitem[\protect\BCAY{{ten Brinke}, Stimson\BCBL{} \BBA{} Carney}{{ten Brinke}
  \BOthers{}}{2014}]{tenBrinke_14}
{ten Brinke}, L., Stimson, D.\BCBL{} \BBA{} Carney, D.~R.
\newblock{}\BBOP{}2014\BBCP{}.
\newblock{}\BBOQ{}{Some evidence for unconscious lie detection}.\BBCQ{}
\newblock{}\Bem{Psychological Science}, \Bem{25}(5), 1098--1105.
\newblock{}(published online Mar 21, 2014)

\bibitem[\protect\BCAY{Tierney}{Tierney}{2014}]{press_NYT_20140323}
Tierney, J.
\newblock{}\BBOP{}2014\BBCP{}.
\newblock{}\Bem{{At airports, a misplaced faith in body language}.}
\newblock{}New York Times (online: March 23, 2014, {\tt
  http://nyti.ms/1eBmrDL}).

\bibitem[\protect\BCAY{Ulrich \BBA{} Miller}{Ulrich \BBA{}
  Miller}{1993}]{Ulrich_Miller_93}
Ulrich, R.\BCBT{} \BBA{} Miller, J.
\newblock{}\BBOP{}1993\BBCP{}.
\newblock{}\BBOQ{}{Information--processing models generating lognormally
  distributed reaction--times}.\BBCQ{}
\newblock{}\Bem{Journal of Mathematical Psychology}, \Bem{37}, 513--525.

\end{thebibliography}
\end{document}